\documentclass[aps,twocolumn,nofootinbib,showpacs,floatfix]{revtex4}
\usepackage{graphics} 
\usepackage{amssymb,amsmath}
\usepackage{amsmath}
\usepackage{psfrag}
\usepackage{graphicx}
\usepackage{subfig}
\usepackage{multirow}
\usepackage{xcolor}
\usepackage{layout}
\usepackage{textcomp}
\usepackage{soul}
\usepackage{psfrag}
\usepackage{graphicx}
\usepackage{dcolumn}
\usepackage{bm}
\usepackage[colorlinks=true,citecolor=black,linkcolor=black,urlcolor=black,filecolor=black]{hyperref}
\usepackage{mathtools}
\usepackage{epstopdf}
\usepackage{color}

\def\beq{\begin{equation}} 
\def\eeq{\end{equation}} 
\def\bea{\begin{eqnarray}} 
\def\eea{\end{eqnarray}}  
\def\bean{\begin{eqnarray*}} 
\def\eean{\end{eqnarray*}} 
\def\dd{\partial}

\def\bk{{\bf k}}  
  
\def\bv{{\bf v}}
\def\br{{\bf r}}

\def\bp{{\mathbf p}}

\def\bse{\begin{subequations}}
\def\ese{\end{subequations}}
\def\bJ{{\mathbf J}}

\def\bw{\mathbf{w}}

\def\lsim{\raise 0.4ex\hbox{$<$}\kern -0.8em\lower 0.62ex\hbox{$\sim$}} 
\def\gsim{\raise 0.4ex\hbox{$>$}\kern -0.7em\lower 0.62ex\hbox{$\sim$}}

\def\bk{\mathbf{k}}

\def\f0N{f_0^{(N)}}
\def\bec{\begin{center}}
\def\eec{\end{center}}

\newcommand{\delf}{\delta\! f}
\newcommand{\fldelf}{\widetilde{\delta\! f}}
\newcommand{\pfpt}{\frac{\partial f}{\partial t}}

\newcommand{\ppv}{\frac{\partial}{\partial \mathbf{v}}}
\newcommand{\pfpr}{\frac{\partial f}{\partial \mathbf{r}}}
\newcommand{\pfpv}{\frac{\partial f}{\partial \mathbf{v}}}
\newcommand{\dif}{\mathop{}\!\mathrm{d}}
\newcommand{\ks}{\kappa^{\star}}
\newcommand{\cpv}{\mathcal{P}\!\!\!\int}
\newcommand{\Om}{\Omega}
\newcommand{\bOm}{\mathbf{\Omega}}
\DeclareMathOperator{\sn}{sn}

\DeclareMathOperator{\dn}{dn}
\DeclareMathOperator{\cn}{cn}
\DeclareMathOperator{\q}{q}
\DeclareMathOperator{\K}{K}
\DeclareMathOperator{\E}{E}
\DeclareMathOperator{\F}{F\!}
\DeclareMathOperator{\Res}{Res}
\DeclarePairedDelimiter\abs{\lvert}{\rvert}%

\begin{document} 
\title{Collisional relaxation in the inhomogeneous Hamiltonian-Mean-Field model: diffusion coefficients}

\author{F. P. C.~Benetti$^{1,2}$ and B.~Marcos$^{2}$} 

\affiliation{$^1$Instituto de F\'{\i}sica, Universidade Federal do Rio Grande do Sul, Brazil\footnote{ Caixa Postal 15051, CEP 91501-970, Porto Alegre, RS, Brazil}}
\affiliation{$^2$Universit\'e C\^ote d'Azur, CNRS UMR 7351, LJAD, France\footnote{Parc Valrose 06108 Nice Cedex 02, France}} 

\begin{abstract}  

Systems of particles with long range interactions present two important processes: first, the formation of out-of-equilibrium quasi-stationary states (QSS), and  the collisional relaxation towards Maxwell-Boltzmann equilibrium in a much longer timescale. In this paper, we study the collisional relaxation in the Hamiltonian-Mean-Field model (HMF) using the appropriate kinetic equations for a system of $N$ particles at order $1/N$ : the Landau equation when collective effects are neglected and the Lenard-Balescu equation when they are taken into account. We derive explicit expressions for the diffusion coefficients using both equations for any magnetization, and we obtain analytic expressions for highly clustered configurations. An important conclusion is that in this system collective effects are crucial in order to describe the relaxation dynamics. We  compare the diffusion calculated with the kinetic equations with simulations set up to simulate the system with or without collective effects, obtaining a very good agreement between theory and simulations.

\end{abstract}    
\pacs{98.80.-k, 05.70.-a, 02.50.-r, 05.40.-a}    
\maketitle   
\date{today}

\section{introduction}

Systems with long range interactions present the generic evolution in
two distinct stages: first, the evolution to a quasi-stationary state
in a process called {\it collisionless (or violent) relaxation} \cite{Lyn1967} in a timescale
$\tau_{dyn}$, and second, the evolution towards thermodynamic
equilibrium in the so-called {\it{collisional relaxation process}}, in
a timescale of order $\tau_{coll}\sim N^\delta \tau_{dyn}$, where
$\delta>0$ depends on the system considered. The mechanism of
collisional relaxation is qualitatively well-known since the seminal
work of Chandrasekhar \cite{chandra_42}: the main elements are
{\it{two-body}} collisions, which randomizes the velocity of the
particles, leading to a Maxwell-Boltzmann velocity distribution.
Using simple calculations and approximating the system as spatially
homogeneous, Chandrasekhar was able to determine that, for gravitational
systems in three dimensions, $\tau_{coll}\sim\tau_{dyn} N/ \ln N
$. This approach was subsequently used by other authors, notably
H\'enon in the sixties (see e.g. \cite{henon_73}) and lead to the development of Fokker-Planck techniques. All these methods share the same
feature of approximating the system as homogeneous.  For example, in
the {\em orbit--averaging} approach (see e.g. \cite{BinTre2008}), diffusion coefficients are
computed approximating the system as homogeneous, and then they are
averaged over the actual orbits of the particles. This method is
used because it is technically difficult to compute diffusion
coefficients for inhomogeneous configurations, essentially because the trajectories of the unperturbed particles (i.e. in the
mean field-limit) would need to be computed, which is generally a very difficult
task. Moreover, using this approach it is not possible to take into
account {\em collective effects}, which can be important for some
systems and configurations, which we will see it is the case in the
present work.

At the same time, a rigorous kinetic theory for (repulsive, neutral) plasmas was
being developed first by Landau (introducing notably the concept of
{\it Landau damping}) and subsequently by other authors such as Lenard, Balescu
etc (see e.g. \cite{balescu_97}). When the system is neutral, the mean field configuration is homogeneous, and it is therefore possible to attack
the problem in an essentially analytical way, including even
collective effects.  

In the last years a rigorous kinetic theory for inhomogeneous
configurations has been developed by different authors \cite{kandrup_81,luciani_87,Hey2010,Cha2013,Cha2012a}. In these works, the general procedure
in order to compute kinetic equations at order $1/N$ has been
described. There are, however, many practical difficulties when trying
to compute quantities of interest such as the diffusion
coefficients, and this for various reasons.  The natural way to write
these equations is to use {\em angle-action} variables (see e.g. \cite{valageasOSC_2}). To compute
them as a function of the natural variables $(x,v)$ is technically
equivalent to solving the equations of motion for the unperturbed
($N\to\infty$) potential, which is in general impossible
analytically. The subsequent calculation of the diffusion coefficient
(which involves e.g. Fourier transform about the angle variable)
becomes (even numerically) very difficult. For this reason we are only aware of
the study of self-gravitating tepid discs 
\cite{FouPic2015a,FouPic2015b}. 
In this case, it is possible to make controlled approximations
which makes the semi--analytical calculations feasible.

In this paper we have chosen to study {\em exactly} a sufficiently simple model in order to compute the diffusion coefficients without approximations (up to order $1/N$). To do so, we use the popular {\em Hamiltonian Mean Field model} (HMF) \cite{antoni_95}, which has widely been used to study long range systems. 
Its simplicity permits to compute some analytical and numerical quantities which would be impossible in more realistic models such as three-dimensional gravity. 
For this reason, the diffusion coefficients have already been studied in the much simpler spatially homogeneous configuration \cite{bouchet_05}. Our work has two main objectives: on one side, it  will permit to compare the diffusion coefficients with numerical simulations in order to check  the validity of the assumptions made deriving the kinetic equations in the case of spatially inhomogeneous distributions. 
On the other side, it will set up the method to solve numerically the Lenard-Balescu equation not only for the HMF but also for other  more complicated models, as self-gravitating systems.

The paper is organized as follows: in the first section we summarize the kinetic theory we will apply in the paper. In the next section we apply the equations for the HMF to compute the diffusion coefficients, giving also analytical results for some cases. Then we compare the theoretical predictions with molecular dynamics simulations, including or not collective effects, and then we give conclusions and perspectives.

\section{Kinetic theory}

The evolution of an $N$-body system under Hamiltonian dynamics can be
described using kinetic theory.  The approach outlined in this section
follows that of several previous works (see Introduction) and is summarized in e.g.\cite{CamDau2009}
\footnote{
	Here, we use the Klimontovich formulism; the same equations
may be obtained from the BBGKY (Bogoliubov-Born-Green-Kirkwood-Yvon) hierarchy, see i.e.~\cite{Hey2010}.
}
.  The problem addressed by this kinetic approach
is the following: given a set of $N$ particles of mass $m$ with
initial positions $\{\br_i\}$ and velocity $\{\bv_i\}$ and their
Hamiltonian equations of motion, how and to what steady state will
they evolve?  We start with the discrete distribution function
$f_d(\br,\bv,t)$, which contains all the information of the state of
the system at a given time $t$, 
\beq
f_d(\br,\bv,t)=m\sum_{i=1}^N\delta[\br-\br_i(t)]\delta[\bv-\bv_i(t)].\label{eq:fdisc}
\eeq 
The evolution of the discrete distribution function is given
exactly by the Klimontovich equation \cite{Cha2012}
\begin{align}
	\frac{\partial f_d}{\partial t}&+\bv\cdot\frac{\dd f_d}{\dd\br}
	-\frac{\dd\phi_d}{\dd\br}\cdot\frac{\dd f_d}{\dd\bv}=0,\label{eq:klim}\\ 
\phi_d(\br,t)&=\int
u(\abs{\br-\br'})f_d(\br',\bv',t)\dif\br'\dif\bv'.\label{eq:phid}
\end{align}
where $\phi_d(\br,t)$ is the discrete convolution potential,
$u(\br-\br')$ is the pair interaction potential between particles at
positions $\br$ and $\br'$
and $\frac{\dd f}{\dd \mathbf{u}}=\sum_{i=1}^d\frac{\dd f}{\dd u_i}\mathbf{e}_i$ and $d$ is the spatial dimension
.

For a given initial distribution
$f_0^d(\br,\bv)=m\sum_{i=1}^N\delta[\br-\br_i(t=0)]\delta[\bv-\bv_i(t=0)]$,
the discrete distribution is determined at all future times $t$. A
smooth distribution function can be obtained by averaging over an
ensemble of initial conditions, 
\beq 
f(\br,\bv,t)=\langle
f_d(\br,\bv,t)\rangle\label{eq:fsmooth} 
\eeq 
and thus $f_d(\br,\bv,t)=f(\br,\bv,t)+\delf(\br,\bv,t)$.

The same smoothing process can be done for the Klimontovich
equation. Since averages over the fluctuations are zero, this leads to
\beq 
\pfpt+\bv\cdot\pfpr
  -\frac{\dd\phi}{\dd\br}\cdot\pfpv
  =\ppv\cdot\langle\delf\frac{\dd\delta
\phi}{\dd\br}\rangle.\label{eq:quasilin1} 
\eeq 
The above equation gives the
evolution of the smooth distribution due to correlation between its
own fluctuations and the fluctuation of the smooth potential
$\phi(\br,t)$, determined by
$\phi_d(\br,t)=\phi(\br,t)+\delta\phi(\br,t)$, where
\begin{align}
\phi(\br,t)&=\int
u(\abs{\br-\br'})f(\br',\bv',t)\dif\br'\dif\bv'\label{eq:phismooth}\\ \delta\phi(\br,t)&=\int
u(\abs{\br-\br'})\delf(\br',\bv',t)\dif\br'\dif\bv'.\label{eq:delphi}
\end{align}

Subtracting equation \eqref{eq:quasilin1} from the Klimontovich
equation and keeping only terms of order lower than
$\mathcal{O}(1/N)$ 
gives the linearised Klimontovich equation,
\beq \frac{\partial \delf}{\partial
  t}+\bv\cdot\frac{\dd\delf}{\dd\br}-\frac{\dd\delta\phi}{\dd\br}\cdot
  \pfpv-\frac{\dd\phi}{\dd\br}\cdot\frac{\dd\delf}{\dd\bv}
  =0\label{eq:quasilin2}.  \eeq 
  The system of equations
\eqref{eq:quasilin1} and \eqref{eq:quasilin2} are known as the
quasi-linear approximation, since in the first equation the correlation
term on the right-hand side is of order $1/N$, while in the second
equation all terms of order $1/N$ or higher have been neglected. 

\subsection{Homogeneous systems}

We will first give a brief derivation of the kinetic equations for the spatially homogeneous case. It is technically simpler than the inhomogeneous one while sharing the same ideas. 
In this case $f=f(\bv,t)$, so equations \eqref{eq:quasilin1} and \eqref{eq:quasilin2} become
\begin{subequations}
\label{kinetic-homo}
\begin{align}
	\pfpt&=\ppv\cdot\langle\delf \frac{\dd\delta\phi}{\dd\br}\rangle,\label{eq:quasilinhom1}\\
	\frac{\partial \delf}{\partial t}&+\bv\cdot\frac{\dd\delf}{\dd\br}-\frac{\dd\delta\phi}{\dd\br}\cdot\pfpv=0.\label{eq:quasilinhom2}
\end{align}
\end{subequations}

The fluctuation terms are more easily dealt with by using the Fourier-Laplace transforms
\beq
\fldelf(\bk,\bv,\omega)=\frac{1}{(2\pi)^d}\int\dif\br\int_0^{\infty}\dif t\, e^{-i(\bk\cdot\br-\omega t)}\delf(\br,\bv,t),\label{eq:ftransform}
\eeq
and
\beq
\widetilde{\delta\phi}(\bk,\omega)=\frac{1}{(2\pi)^d}\int\dif\br\int_0^{\infty}\dif t\, e^{-i(\bk\cdot\br-\omega t)}\delta\phi(\br,t).\label{eq:phitransform}
\eeq

Taking the Fourier-Laplace transform of equation \eqref{eq:quasilinhom2}, we have
\beq
\widehat{\delta f}({\bf k},{\bf v},0)-i(\bk\cdot\bv-\omega)\, \fldelf({\bf k},{\bf v},\omega)+i {\bf k}\cdot\pfpv\, \widetilde{\delta\phi}({\bf k},\omega)=0,
\eeq
where
\beq
\widehat{\delta\! f}({\bf k},{\bf v},0)=\int\frac{\dif{\bf r}}{(2\pi)^d}\, e^{-i{\bf k}\cdot {\bf r}}\delf({\bf r},{\bf v},0).
\eeq

From the above equation, we can isolate $\fldelf$ and thus find an expression relating the fluctuations of the distribution function and the fluctuations of the potential and the initial condition,
\beq
\label{coll-ini}
\widetilde{\delta\! f}=\underbrace{\frac{\bk\cdot\frac{\partial f}{\partial \bv}\widetilde{\delta\phi}(\bk)}{\bk\cdot\bv-\omega}}_{\mathclap{\substack{\text{collective}\\\text{effects}}}}
+\underbrace{\frac{\widehat{\delta\! f}(\bk,\bv,0)}{i(\bk\cdot\bv-\omega)}}_{\mathclap{\substack{\text{initial}\\\text{conditions}}}}.
\eeq
Because collective effects are difficult to compute
analytically, a common approximation found in the literature consists in neglecting them
(see e.g. ~\cite{Cha2013}). In this paper we will consider the complete problem, and we will study their importance in the inhomogeneous HMF. 

The next step in the derivation consists in
expressing the Fourier transform of the fluctuation of the potential
$ \widetilde{\delta\phi}(\bk,\omega)$ as a function of the
fluctuation $\fldelf(\bk,w)$. To do so, we integrate
equation~\eqref{coll-ini} over $\bv$, and using the Fourier transform of
equation~\eqref{eq:delphi}, we get
\beq
\label{fluctuat}
\int_{-\infty}^\infty \dif\bv \fldelf(\bk,\bv,\omega) = \frac{1}{\epsilon(\bk,\omega)}
\int_{-\infty}^\infty \dif\bv \frac{  \widehat{\delta\! f}(\bk,\bv,0)}{i(\bv\cdot\bk-\omega)},
\eeq
where we have defined the plasma response dielectric function
\beq
\label{epsilon-definition}
\epsilon(\bk,\omega) = 1 - \hat u(\bk)\int \dif\bv \frac{\bk\cdot \dd f(\bv)/\dd\bv}{\bv\cdot\bk-\omega}.
\eeq
Using again equations~\eqref{eq:delphi} and~\eqref{fluctuat}, we get
\begin{eqnarray}
\label{fluctu-phi}
\widetilde{\delta\phi}(\bk,\omega)&=&\hat u(\bk)\int_{-\infty}^\infty \dif\bv \fldelf(\bk,\bv,\omega)\\\nonumber
&=&\frac{\hat u(\bk)}{\epsilon(\bk,\omega)}
\int_{-\infty}^\infty \dif\bv \frac{\widehat{\delta\! f}(\bk,\bv,0)}{i(\bp\cdot\bk-\omega)}
\end{eqnarray}
Inserting equations~\eqref{coll-ini} and \eqref{fluctu-phi} in equation~\eqref{eq:quasilinhom1}, after some algebra, we get the
Lenard-Balescu equation (using the notation \cite{Cha2012}):

\begin{align}\label{eq:LBhom}
\pfpt=&\pi(2\pi)^dm\sum_{i,j=1}^d\frac{\partial}{\partial v_i}
\int\dif\bk\dif\bv'k_ik_j\frac{\hat{u}(\bk)^2}{\abs{\epsilon(\bk,\bk\!\cdot\!\bv)}^2}\nonumber\\
&\times\!\delta\!\left[\bk\!\cdot\!(\bv-\bv')\right]\!
\left(\frac{\partial}{\partial v_j}-\frac{\partial}{\partial v_j'}\right)\!f(\bv,t)f(\bv',t)
\end{align}
When collective effects are neglected, i.e., the first term of equation~\eqref{coll-ini} is neglected, it is simple to see from equation~\eqref{epsilon-definition}, that $\epsilon(\bk,\omega)=1$.

\subsection{Inhomogeneous systems}

In inhomogeneous systems, the strategy is to use, instead of the variables $(\br,\bv)$, the angle-action variables $(\bw,\bJ)$ corresponding to the Hamiltonian $\mathcal H$ of smooth dynamics (i.e. the one corresponding to the limit $N\to\infty$)~\cite{BinTre2008}. Using these variables, particles described by the Hamiltonian $\mathcal H$ keep their action $\bJ$ constant during the dynamic and their angle evolves with time as $\mathbf{w}=\mathbf{\Omega}(\mathbf{J})t+\mathbf{w}_0$ where $\mathbf{w}_0$ is the angle at $t=0$ and $\mathbf{\Omega}(\mathbf{J})=\partial\mathcal{H}/\partial\mathbf{J}$ is the angular frequency \cite{goldstein_02}. The system thus
becomes ``homogeneous'' in the new coordinates~\cite{lichtenberg2010regular}.

The equations for evolution of smooth distribution function $f$ and its fluctuation $\delf$ are~\cite{luciani_87,Cha2012a}
\begin{subequations}
\label{eq:aaevolve}
\begin{align}
\frac{\partial f(\mathbf{J})}{\partial t}+\left[\mathcal{H}(\mathbf{J}),f(\mathbf{J})\right]=-\left\langle\left[\delta\phi,\delf(\mathbf{J})\right]\right\rangle,\label{eq:aaevolve1}\\
\frac{\partial\delta\! f(\mathbf{J})}{\partial t}+\left[\mathcal{H}(\mathbf{J}),\delf(\mathbf{J})\right]+\left[\delta\phi,f(\mathbf{J})\right]=0,\label{eq:aaevolve2}
\end{align}
\end{subequations}
where $\phi$ is the smooth mean-field potential and $\delta\phi$ is its fluctuation, and $[\mathcal{H},B]=\frac{\partial \mathcal{H}}{\partial \mathbf J}\frac{\partial B}{\partial \mathbf{w}}
-\frac{\partial\mathcal{H}}{\partial\mathbf{w}}\frac{\partial B}{\partial \mathbf{J}}$ are Poisson brackets with action-angle variables as the canonical coordinates.

Since by construction $\partial\mathcal{H}/\partial\mathbf{w}=0$ and $\partial f/\partial\mathbf{w}=0$, the terms in Poisson brackets reduce to
\begin{align}
\left[\mathcal{H},\delta\! f\right]=\frac{\partial\mathcal{H}}{\partial\mathbf{J}}\frac{\partial\delta\! f}{\partial\mathbf{w}}
=\mathbf{\Omega}(\mathbf{J})\cdot\frac{\partial\delta\! f}{\partial\mathbf{w}},\\
\left[\delta\phi,f\right]=-\frac{\partial\delta\phi}{\partial\mathbf{w}}\cdot\frac{\partial f}{\partial\mathbf{J}}.
\end{align}
Substituting the above in equations \eqref{eq:aaevolve} and averaging over angles $\mathbf{w}$,
\begin{subequations}
\label{kinetic-aa}
\begin{align}
\label{kinetic-aa1}\frac{\dd\overline{f}}{\dd t}=\frac{\partial}{\partial\mathbf{J}}\cdot\left\langle\,\overline{\delta\! f\frac{\partial\delta\phi}{\partial\mathbf{w}}}\,\right\rangle,\\
\label{kinetic-aa2}
\frac{\partial\overline{\delta\! f}}{\partial t}+\mathbf{\Omega}(\mathbf{J})\cdot\frac{\partial\overline{\delta\! f}}{\partial\mathbf{w}}-\overline{\frac{\partial\delta\phi}{\partial\mathbf{w}}
\cdot\frac{\partial f}{\partial\mathbf{J}}}=0,
\end{align}
\end{subequations}
where $\overline{A}$ represents the angle-averaging of $A$. From now on, we disregard this notation and write $\overline{A}=A$ for simplicity, but we emphasize that the equations from this point further correspond to the angle-averaged quantities.

Observe that equations~\eqref{kinetic-aa} have the same structure as their homogeneous counterpart equations~\eqref{kinetic-homo} identifying the action $\bJ$ with the velocity $\bv$ and the angle $\bw$ with the spatial variable $\br$. The only difference appears in the second term of equation~\eqref{kinetic-aa2} in which the velocity $\bv$ is substituted by the frequency of the unperturbed orbit $\Omega(\bJ)$. Following then the same procedure than the one described in the homogeneous case, we get the Lenard-Balescu-type kinetic equation (with collective effects) in action-angle variables \cite{Hey2010,Cha2012a},
\begin{align}\label{eq:LBaa}
\pfpt=\pi(2\pi)^d&m\frac{\partial}{\partial\bJ}\cdot\sum_{\bk,\bk'}
\int d\bJ'\bk\frac{\delta[\bk\!\cdot\!\bOm(\bJ)-\bk'\!\cdot\!\bOm(\bJ')]}{|D_{\bk,\bk'}(\bJ,\bJ',\bk\cdot\bOm(\bJ))|^2}\nonumber\\
&\times\!
\left(\bk\!\cdot\!\frac{\partial}{\partial\mathbf{J}}-\bk'\!\cdot\!\frac{\partial}{\partial\mathbf{J}'}\right)
f(\mathbf{J},t)f(\mathbf{J}',t)
\end{align}
where
\beq\label{eq:D}
\frac{1}{D_{\bk,\bk'}(\mathbf{J},\mathbf{J}',\omega)}=
\sum_{\alpha,\alpha'}\hat{\Phi}_{\alpha}(\bk,\mathbf{J})(\epsilon^{-1})_{\alpha,\alpha'}(\omega)
\hat{\Phi}^{\star}_{\alpha'}(\bk',\mathbf{J}'),
\eeq
and $\epsilon_{\alpha\alpha'}(\omega)$ is the dielectric tensor
\begin{align}\label{eq:eps}
\epsilon_{\alpha\alpha'}(\omega)=\delta_{\alpha\alpha'}+(2\pi)^d\sum_{\bk}&\int d\mathbf{J}
\frac{\bk\cdot\partial f/\partial\mathbf{J}}{\bk\cdot\mathbf{\Omega}(\bJ)-\omega}\nonumber\\
&\times\hat{\Phi}^{\star}_{\alpha}(\bk,\mathbf{J})\hat{\Phi}_{\alpha'}(\bk,\mathbf{J}).
\end{align}

The indices $(\alpha,\alpha')$ are labels for the bi-orthogonal basis $\{\rho_{\alpha},\Phi_{\alpha}\}$, where
$\rho(\br)=\int f(\br,\bv,t)\dif\bv$, which satisfies~\cite{Kal1976}
\begin{align}
\int u(|\br-\br'|)\rho_{\alpha}(\br')\dif\br'=\Phi_{\alpha}\\
\int\rho_{\alpha}(\br)\Phi_{\alpha'}^{\star}(\br)\dif\br=-\delta_{\alpha,\alpha'}.
\end{align}
The terms $\hat{\Phi}_{\alpha}$ are the Fourier transforms of the potential in the bi-orthogonal representation
with respect to the angles,
\beq
\hat{\Phi}_{\alpha}(\bk,\mathbf{J})=\frac{1}{(2\pi)^d}\int \dif\mathbf{w}e^{-i\bk\cdot\mathbf{w}}
\Phi_{\alpha}(\mathbf{w},\mathbf{J}).
\eeq
The Lenard-Balescu equation \eqref{eq:LBaa} gives the evolution of
$f$ due to the inclusion of a finite-$N$ correction to the
collisionless (Vlasov) kinetic equation. From equation
\eqref{eq:LBaa}, we see that the evolution, which slowly deforms the orbits
of constant $\bJ$, is driven by resonances between
orbital frequencies, $\bk\!\cdot\!\bOm(\bJ)=\bk'\!\cdot\!\bOm(\bJ')$. This differs from the homogeneous case, equation \eqref{eq:LBhom}, where $f$
evolves due to the resonances $\bv=\bv'$.

Using the chain rule, the Lenard-Balescu-type equation \eqref{eq:LBaa} can be written in the form of a Fokker-Planck equation
\beq\label{eq:fp}
\pfpt=\sum_{i,j=1}^d\frac{\partial^2 }{\partial J_i\partial J_j}D_{dif}^{ij}(\bJ,t) f(\bJ,t)-\frac{\partial}{\partial \bJ}\cdot \mathbf{D}_{fr}(\bJ,t)f(\bJ,t)
\eeq
where
\begin{align}\label{eq:CEdiff}
D_{dif}^{ij}(\bJ,t)=&\pi(2\pi)^dm\sum_{\bk,\bk'}\int\dif\bJ'k_i k_j\frac{1}{|D_{\bk,\bk'}(\bJ,\bJ',\bk'\cdot\bOm(\bJ'))|^2}\nonumber\\
&\times\delta[\bk\cdot\bOm(\bJ)-\bk'\cdot\bOm(\bJ')]
f(\bJ',t)
\end{align}
is the diffusion coefficient and the friction coefficient is 
\begin{align}\label{eq:CEfric1}
\mathbf{D}_{fr}(\bJ,t)=&
\pi(2\pi)^dm\! \sum_{\bk,\bk'}\!
\int\!\dif\bJ'f(\bJ')\,\bk\left(\bk\frac{\dd}{\dd\bJ}-\bk'\frac{\dd}{\dd\bJ'}\right)\nonumber\\
&\times\frac{\delta[\bk\!\cdot\!\bOm(\bJ)-\bk'\!\cdot\!\bOm(\bJ')]}{|D_{\bk,\bk'}(\bJ,\bJ',\bk'\!\cdot\!\bOm(\bJ'))|^2}
\end{align}
The $i$th component of the friction coefficient \eqref{eq:CEfric1} can
also be written as the sum of the derivative of the
diffusion coefficient, plus a polarization force~\cite{Cha2012a}
\beq\label{eq:dfric}
D_{fr}^i(\bJ,t)=\frac{\partial}{\partial J_i}D_{dif}^{ij}(\bJ,t)+D_{pol}^i(\bJ,t)
\eeq
where the $i$-component of the polarization force is
\begin{align}
D_{pol}^i(\bJ,t)=&\pi(2\pi)^dm\!\sum_{\bk,\bk'}\!\int\dif\bJ'k_i\bk'\frac{1}{|D_{\bk,\bk'}(\bJ,\bJ',\bk'\!\cdot\!\bOm(\bJ'))|^2}\nonumber\\
&\times\delta[\bk\cdot\bOm(\bJ)-\bk'\cdot\bOm(\bJ')]
\frac{\partial f(\bJ',t)}{\partial\bJ'}.
\end{align}
When collective effects are not considered, we have
\beq
\label{cond-nocoll}
\epsilon_{\alpha\alpha'}=\delta_{\alpha\alpha'},
\eeq
and therefore the Landau equation is obtained using
the \emph{bare}, undressed Fourier transforms of the potential,
\beq
\frac{1}{|D_{\bk,\bk'}^{bare}(\bJ,\bJ',\bk'\cdot\bOm(\bJ'))|^2}=|\hat{\Phi}_{\alpha}(\bk,\mathbf{J}) \hat{\Phi}^{\star}_{\alpha}(\bk',\mathbf{J}')|^2.
\eeq

\section{Kinetic equations for the Hamiltonian Mean-Field model}

We will compute explicitly the diffusion coefficients for the HMF model. It is given by the Hamiltonian 
\beq\label{eq:hmfhamiltonian}
H=\sum_{i=1}^N\frac{p^2}{2}-\frac{1}{2N}\sum_{i,j=1}^N\cos(\theta_i-\theta_j).
\eeq
The energy of one particle can be written as
\beq\label{eq:hmfparticleenergy}
h(\theta,p)=\frac{p^2}{2}+\phi(\theta)=\frac{p^2}{2}-\frac{1}{N}\sum_{i=1}^N\cos(\theta_i-\theta).
\eeq
The potential $\phi(\theta)=-1/N\sum_i\cos(\theta_i-\theta)$ can be rewritten as
\begin{align}
\phi(\theta)&=-\frac{\sum_{i=1}^N\cos\theta_i}{N}\cos\theta-\frac{\sum_{i=1}^N\sin\theta_i}{N}\sin\theta\nonumber\\
&=-M_x\cos\theta-M_y\sin\theta
\end{align}
where $\mathbf{M}=(M_x,M_y)$ is the magnetization vector. Its modulus quantifies
how bunched, or clustered, the particles are.
Shifting all angles by a phase $\alpha=\arctan(M_y/M_x)$, we can write the potential simply as a function of the modulus of the magnetization $M$,
\beq\label{eq:phiphase}
\phi(\theta^{\star})=-M\cos\theta^{\star}
\eeq
where $\theta^{\star}=\theta-\alpha$ and $M=M_x=\sum_{i=1}^N\cos\theta^{\star}_i$.
For simplicity, henceforth we denote $\theta^{\star}$ as $\theta$.

\begin{figure}
\begin{centering}
\includegraphics[width=7cm]{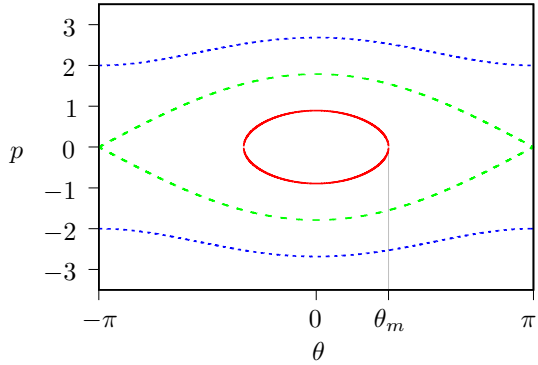}
\caption{Examples of a librating orbit (red solid line), for which $\kappa<1$, a rotating orbit (blue dotted line), for which $\kappa>1$, and the separatrix orbit (green dashed line), for which $\kappa=1$. For the librating orbit, $\theta_m=\arccos(1-2\kappa^2)$, while for the other orbits $\theta_m=\pi$.\label{fig:hmforbits}}
\end{centering}
\end{figure}

\subsection{Action-angle variables}

\begin{figure}
\begin{centering}
\includegraphics[width=8.5cm]{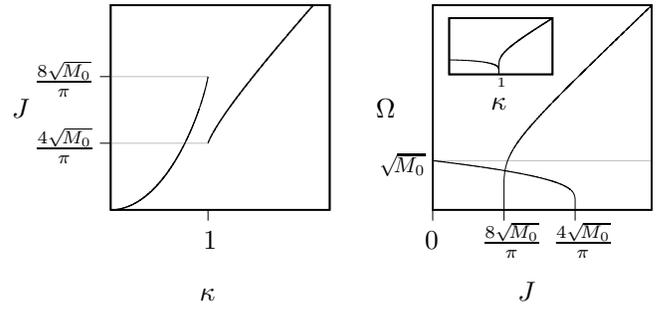}
\caption{Action as a function of $\kappa$ for the HMF model (left), and frequency $\Om$ versus $J$  (inset: $\Om$ vs $\kappa$) (right) for the HMF model.\label{fig:hmfactom}}
\end{centering}
\end{figure}

Inhomogeneous states of the HMF model have previously been studied using action-angle variables in the case of Vlasov stability~\cite{BarOli2010} 
and \cite{Oga2013}. We define our action angle variables in the same way as these references. The action $J$ is defined as
\beq
J=\frac{1}{2\pi}\oint p\dif\theta\nonumber\\
\eeq
with $p=\sqrt{2(h-\phi(\theta))}$, where energy $h$ is the one-particle energy and $\phi(\theta)$ is the mean-field potential, equation \eqref{eq:phiphase}.
The potential can be fully specified with a single scalar quantity, the modulus of the magnetization $M$. 
It is possible to write simply and in a generic way an expression for the action which depends only on the energy of the particle $h$ and the adiabatic, static magnetization $M_0$ (see Appendix \ref{ap:aa})
\beq\label{eq:hmfaction}
J(\kappa)=\frac{4\sqrt{M_0}}{\pi}\begin{cases}
2\left[\E(\kappa)-(1-\kappa^2)\K(\kappa)\right],&\quad\kappa<1\\
\kappa \E\left(\frac{1}{\kappa}\right),&\quad\kappa>1
\end{cases}
\eeq
where
\beq\label{eq:hmfkappa}
\kappa=\sqrt{\frac{h+M_0}{2M_0}}.
\eeq
The action $J$ is discontinuous at the separatrix $\kappa=1$, the boundary between rotating and librating orbits (see Figure \ref{fig:hmforbits}). Figure \ref{fig:hmfactom} shows the action as a function of $\kappa$ and the discontinuity at the separatrix.
 
The frequency $\Om(J)$  is $\Om(J)=\partial h/\partial J$. 
Due to the frequency being non-injective in $J$, and $J$ being a function of elliptical integrals of $\kappa$, it is easier 
to treat all expressions directly as a function of $\kappa$. We use the Jacobian $\partial \kappa/\partial J$ to change variables, 
\beq\label{eq:hmfjacob}
\left[\frac{\partial J}{\partial \kappa}\right]=\frac{4\sqrt{M_0}}{\pi}
\begin{cases}
2\kappa \K(\kappa),&\quad\kappa<1\\
\K\left(\frac{1}{\kappa}\right),&\quad\kappa>1.
\end{cases}
\eeq
Thus the frequency is given by $\Om(J)=(\partial\kappa/\partial J)(\partial h/\partial \kappa)$,
\beq\label{eq:hmfomega}
\Om(\kappa)=\pi\sqrt{M_0}
\begin{cases}
\frac{1}{2\K(\kappa)},&\quad\kappa<1\\
\frac{\kappa}{\K\left(\frac{1}{\kappa}\right)},&\quad\kappa>1.
\end{cases}
\eeq
The explicit expressions for the action-angle variables is a great advantage of the HMF model for the investigating inhomogeneous states. 
For most systems, this is not possible; a few exceptions in astrophysics being spherical potentials and flat axisymmetric potentials such as razor-thin and tepid discs, as well as some nonaxisymmetric potentials such as St\"ackel potentials~\cite{BinTre2008}.

\subsection{Kinetic equations}

For the HMF model, the pair potential $u(\theta-\theta')=-\cos(\theta-\theta')$ can be written in the two-dimensional bi-orthogonal representation as 
$\Phi_c=-\cos[\theta(w,\kappa)]$ and $\Phi_s=-\sin[\theta(w,\kappa)]$,
and its Fourier transforms are
\beq\label{eq:hmfft1}
\begin{aligned}
\hat{\Phi}_c(m,\kappa)=-c_m(\kappa)=\frac{-1}{2\pi}\int_{-\pi}^{\pi}\cos[\theta(w,\kappa)]e^{-imw}dw,\\
\hat{\Phi}_s(m,\kappa)=-s_m(\kappa)=\frac{-1}{2\pi}\int_{-\pi}^{\pi}\sin[\theta(w,\kappa)]e^{-imw}dw.
\end{aligned}
\eeq
These can be written more simply as (see Appendix~\ref{ap:ellip})
\beq\label{eq:hmfcm}
c_n(\kappa)=\begin{cases}
\frac{\pi^2}{\K(\kappa)^2}\frac{\abs{n}\q(\kappa)^{\abs{n}/2}}{1-\q(\kappa)^{\abs{n}}}&\quad \kappa<1,\;n\text{ even},\\
0&\quad \kappa<1,\; n\text{ odd},\\
\frac{2\pi^2\kappa^2}{\K\left(\frac{1}{\kappa}\right)^2}\frac{\abs{n}\q\left(\frac{1}{\kappa}\right)^{\abs{n}}}
{1-\q\left(\frac{1}{\kappa}\right)^{2\abs{n}}}&\quad\kappa>1,
\end{cases}
\eeq
and
\beq\label{eq:hmfsm}
s_n(\kappa)=\begin{cases}
0&\quad\kappa<1,\;n\text{ even},\\
-i\frac{\pi^2}{\K(\kappa)^2}\frac{n \q(\kappa)^{\abs{n}/2}}{1+\q(\kappa)^{\abs{n}}}&\quad\kappa<1,\;n\text{ odd},\\
-i\frac{2\pi^2\kappa^2}{\K\left(\frac{1}{\kappa}\right)^2}\frac{n \q\left(\frac{1}{\kappa}\right)^{\abs{n}}}
{1+\q\left(\frac{1}{\kappa}\right)^{2\abs{n}}}&\quad\kappa>1,\;p>0\\
i\frac{2\pi^2\kappa^2}{\K\left(\frac{1}{\kappa}\right)^2}\frac{n \q\left(\frac{1}{\kappa}\right)^{\abs{n}}}
{1+\q\left(\frac{1}{\kappa}\right)^{2\abs{n}}}&\quad\kappa>1,\;p<0
\end{cases}
\eeq
where $\q(k)=\exp[-\pi\K(\sqrt{1-k^2})/\K(k)]$. To switch variables from $J$ to $\kappa$, we use the Dirac delta identity $\delta[f(x)]=\sum_{x^*}\delta(x-x^*)/|\partial f/\partial x|_{x^*}$ (where $x^*$ are the roots of $f(x)$).
Thus, the Lenard-Balescu equation for the HMF model  is
\begin{align}\label{eq:hmflb}
&\pfpt=\frac{2\pi^2}{N}\left|\frac{\partial J}{\partial \kappa}\right|^{-1}\!\frac{\partial}{\partial \kappa}\!
\sum_{n,n'=-\infty}^{\infty}\!\int
\frac{\dif\kappa'n|\partial J'/\partial \kappa'|}{|D_{nn'}(\kappa,\kappa',n\Omega(\kappa))|^2}\nonumber\\
&\!\times\!\sum_{\ks}\!\frac{\delta(\kappa'\!-\!\ks)}{|n'\frac{\partial \Omega}{\partial\kappa'}|_{\ks}}
\!\left(\!n\!\left|\frac{\partial J}{\partial \kappa}\right|^{-1}\!\!\!\frac{\partial }{\partial\kappa}
\!-\!n'\left|\frac{\partial J'}{\partial \kappa'}\right|^{-1}\!\!\!\frac{\partial}{\partial\kappa'}\!\right)\!
f(\kappa,t)f(\kappa',t),
\end{align}
where $\ks$ are the roots of the equation $m\Omega(\kappa)-m'\Omega(\kappa')=0$, the Jacobian 
$|\partial J/\partial \kappa|$ is given by equation \eqref{eq:hmfjacob}, and $\partial\Om/\partial\kappa$ is
\beq\label{eq:domdk}
\frac{\partial\Om}{\partial\kappa}=
\pi\sqrt{M_0}\begin{cases}
\begin{aligned}
\frac{\E(\kappa)+(\kappa^2-1)\K(\kappa)}{2\kappa(\kappa^2-1)\K^2(\kappa)},\qquad&\kappa<1,\\
\frac{\kappa^2\E\left(\frac{1}{\kappa}\right)}{(\kappa^2-1)\K^2\left(\frac{1}{\kappa}\right)},
\qquad&\kappa>1.
\end{aligned}
\end{cases}
\eeq
The associated diffusion coefficient is
\beq\label{eq:hmfCEdif}
D_{dif}(\kappa)=\frac{2\pi^2}{N}\!\!\sum_{n,n'=\infty}^{\infty}
\sum_{\ks}
\frac{n^2|\partial J/\partial \kappa|_{\ks}}{\left|D_{nn'}(\kappa,\ks,n\Om(\kappa))\right|^2}
\frac{f(\ks,t)}{\left|n'\frac{\partial\Om}{\partial\kappa'}\right|_{\ks}}
\eeq
and the polarization coefficient is
\beq\label{eq:hmfCEpol}
D_{pol}(\kappa)=\frac{2\pi^2}{N}\!\!\sum_{n,n'=-\infty}^{\infty}\sum_{\ks}\frac{n\,n'}{\abs{D_{nn'}(\kappa,\ks,n\Om(\kappa))}^2}
\frac{\partial f/\partial \kappa'|_{\ks}}{\left|n'\frac{\partial \Om}{\partial \kappa'}\right|_{\ks}}.
\eeq
Equation \eqref{eq:D}, which determines $D_{nn'}(\kappa,\kappa',\omega)$, becomes
\beq\label{eq:hmfD}
\frac{1}{D_{nn'}(\kappa,\kappa',\omega)}=\frac{c_n(\kappa)c_{n'}(\kappa')}{\epsilon_{cc}(\omega)}
-\frac{s_n(\kappa)s_{n'}(\kappa')}{\epsilon_{ss}(\omega)}.
\eeq
If collective effects are neglected, $\epsilon_{cc}=\epsilon_{ss}=1$, and we get simply
\beq\label{eq:hmfDbare}
\frac{1}{D^{bare}_{nn'}(\kappa,\kappa')}=c_n(\kappa)c_{n'}(\kappa')-s_n(\kappa)s_{n'}(\kappa').
\eeq
If collective effects are not neglected, it is necessary to compute numerically the dielectric tensor, with the procedure we detail below.

\subsection{Numerical computation of the dielectric tensor}

The $cc$ and $ss$ components of the dielectric tensor are
\beq\label{eq:hmfepscc}
\epsilon_{cc}(\omega)=1+2\pi\sum_{\ell=-\infty}^{\infty}\int_0^\infty
\dif\kappa\frac{g_\ell^{cc}(\kappa)}{\Om(\kappa)-\omega/\ell},
\eeq
and
\beq\label{eq:hmfepsss}
\epsilon_{ss}(\omega)=1+2\pi\sum_{\ell=-\infty}^{\infty}\int_0^\infty
\dif\kappa\frac{g_\ell^{ss}(\kappa)}{\Om(\kappa)-\omega/\ell},
\eeq
respectively, where to simplify the notation we have defined
\bse
\begin{align}
g_\ell^{cc}(\kappa)&=\abs{c_\ell(\kappa)}^2\partial f/\partial\kappa\\ 
g_\ell^{ss}(\kappa)&=\abs{s_\ell(\kappa)}^2\partial f/\partial\kappa.
\end{align}
\ese
 The off-diagonal terms, involving products of the type $c_n(\kappa)s_{n'}(\kappa')$, are zero after integration. 

\begin{figure}
\begin{centering}
\includegraphics[width=8cm]{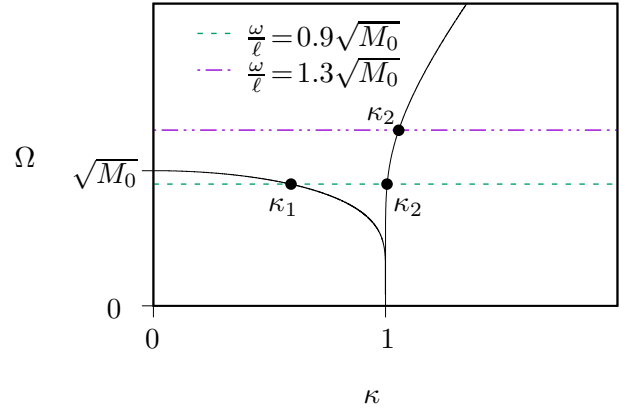}
\caption{Poles of integral in the dielectric tensor components \eqref{eq:hmfepscc} and \eqref{eq:hmfepsss}. For $\omega/\ell>\Omega_0$ (dash-dotted line), 
only one pole occurs ($\kappa_2$), while for $0<\omega/\ell<\Omega_0$ (dashed line), there are two ($\kappa_1$ and $\kappa_2$).
\label{fig:hmfpoles}}
\end{centering}
\end{figure}

The integrals in equations \eqref{eq:hmfepscc} and \eqref{eq:hmfepsss} must be performed carefully due to the poles at $\omega=\ell\Omega(\kappa)$. 
Poles can only occur if $\ell$ and $\omega$ are of the same sign. 
Moreover, the number of poles depends on the value of $\omega$, since $\Omega(\kappa)$ can have the same value at two different values of $\kappa$ for $\Omega(\kappa)<\Omega_0$ where $\Om_0=\Om(0)=\sqrt{M_0}$. 
Therefore, we distinguish between the following cases (see Figure \ref{fig:hmfpoles}).
\begin{enumerate}
\item $\omega/\ell<0$: no poles;
\item $0<\omega/\ell<\Omega_0$: one pole $\kappa_1<1$ and one pole at $\kappa_2>1$;
\item $\omega/\ell>\Omega_0$: one pole at $\kappa_2>1$.
\end{enumerate}

For each case, the integrals must be separated into different regions. In all cases we separate between the regions $\kappa\in(0,1)$ and $\kappa\in(1,\infty)$, due to the different expressions of $\Omega(\kappa)$, $c_n(\kappa)$ and $s_n(\kappa)$ in the two domains. Therefore, for case $1$, the integrals in equations \eqref{eq:hmfepscc}, \eqref{eq:hmfepsss} is
\begin{align}
\int\dif\kappa\frac{g_\ell^{cc/ss}(\kappa)}{\Om(\kappa)-\omega/\ell}=&\int_0^{1}\!\dif\kappa
\frac{g_\ell^{cc/ss}(\kappa)}{\Om(\kappa)-\omega/\ell}\nonumber\\
&+\int_{1}^\infty\dif\kappa
\frac{g_\ell^{cc/ss}(\kappa)}{\Om(\kappa)-\omega/\ell}.
\end{align}
For case $2$, we must use the Landau contour in both regions,
\begin{align}
\int\dif\kappa\frac{g_\ell^{cc/ss}(\kappa)}{\Om(\kappa)-\omega/\ell}=&
\cpv_0^{1}\!\!\!\dif\kappa\frac{g_\ell^{cc/ss}(\kappa)}{\Om(\kappa)-\omega/\ell}
+i\pi\Res{\kappa_1}\nonumber\\
&+\cpv_{1}^{\infty}\!\!\!\dif\kappa\frac{g_\ell^{cc/ss}(\kappa)}{\Om(\kappa)-\omega/\ell}+i\pi\Res{\kappa_2},
\end{align}
and for case $3$, only in the second region,
\begin{align}
\int\dif\kappa\frac{g_\ell^{cc/ss}(\kappa)}{\Om(\kappa)-\omega/\ell}=&
\int_0^{1}\!\dif\kappa\frac{g_\ell^{cc/ss}(\kappa)}{\Om(\kappa)-\omega/\ell}\nonumber\\
&+\cpv_{1}^{\infty}\!\!\!\dif\kappa\frac{g_\ell^{cc/ss}(\kappa)}{\Om(\kappa)-\omega/\ell}+i\pi\Res{\kappa_2}
\end{align}
where $\mathcal P\int$ denotes the Cauchy principal value and $\Res{x}$ is the residue of the integrand at $x$. 

Equations \eqref{eq:hmfCEdif}, \eqref{eq:hmfD}, \eqref{eq:hmfepscc}, \eqref{eq:hmfepsss}, with $\Omega(\kappa)$, $s_m(\kappa)$ and $c_m(\kappa)$ determined by equations \eqref{eq:hmfomega}, \eqref{eq:hmfcm}, and \eqref{eq:hmfsm}, respectively, enable us to calculate the diffusion coefficient of the HMF model in action-angle variables, with collective effects.
The same can be done neglecting collective effects, using the same equations with $\epsilon_{cc}=\epsilon_{ss}=1$. The inclusion or exclusion of collective effects greatly affects the resulting diffusion coefficient. 
This is shown in Figure \ref{fig:hmfCENCEcompare}, where we present diffusion coefficients 
considering a thermal bath, 
\beq
\label{gaussian-def}
f(\kappa,t)=C\exp[-\beta M_0(2\kappa^2-1)],
\eeq
for two equilibrium 
configurations $(\beta,M_0)$, where $C=\sqrt{\beta/(2\pi)^3}/I_0(\beta M_0)$ and $I_n(z)$ is the $n$th-order modified Bessel function of the first kind. For the numerical results, all sums over $n$, $n'$ and $\ell$ are truncated at $n_{max}=6$ and $\ell_{max}=6$ respectively (although normally $n_{max}=4$ and $\ell_{max}=2$ suffice).

\begin{figure}
\begin{centering}
\includegraphics[width=8.5cm]{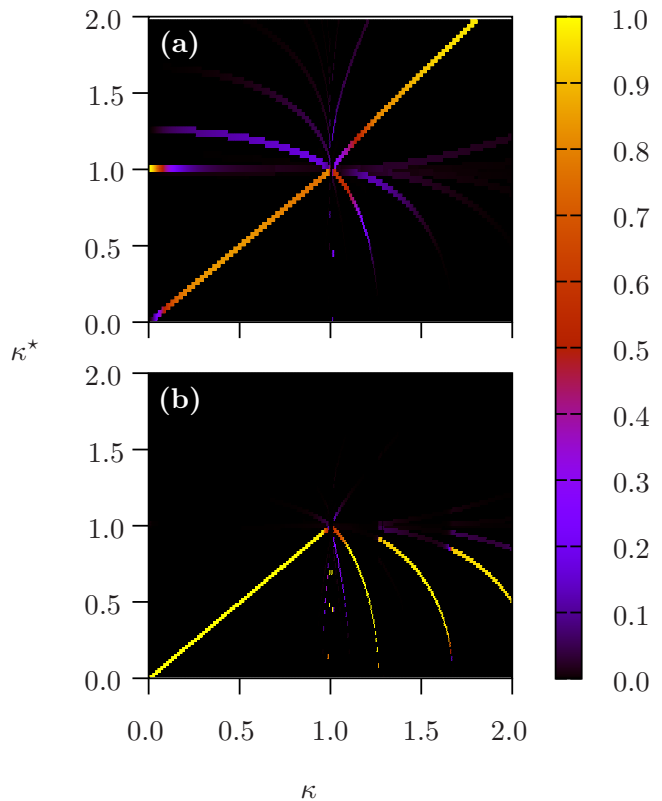}
\caption{Normalized contribution to the Lenard-Balescu diffusion coefficient $D_{dif}(\kappa)$, equation \eqref{eq:hmfCEdif}, as a function of $\kappa^{\star}$. Both panels correspond to thermal equilibrium distributions, but with different magnetizations: (a) is almost homogeneous, $M_0=0.05$, and (b) is highly inhomogeneous, $M_0=0.9$. In the latter case, most of the contribution comes from resonances at $\ks<1.0$ (below the separatrix), while for the nearly homogeneous system this is not the case.
\label{fig:hmfcontribs}}
\end{centering}
\end{figure}

From the forms of equations of the diffusion coefficients \eqref{eq:hmfCEdif} we see that the 
contributions to the diffusion of a particle with a parameter $\kappa$ come from its resonances with particles of parameter $\ks$, where $\ks$ and $\kappa$ satisfy $n\Omega(\kappa)=n'\Omega(\ks)$ and $n,n'$ are integers. In order to see how each resonance contributes to the diffusion coefficient, in Figure \ref{fig:hmfcontribs} we plot maps showing the
normalized contribution of each term in the $\ks$ sum, for a given $\kappa$, for a thermal distribution function corresponding to $M_0=0.05$ (top) and $M_0=0.9$ (bottom). In other words, if we write the diffusion coefficient as 
\beq
D_{dif}(\kappa)=\sum_{\ks} \gamma(\kappa,\ks),
\eeq
the color map shows $\gamma(\kappa,\ks)/D_{dif}(\kappa)$.

In the highly inhomogeneous case, $M_0=0.9$, almost all the contribution comes from $\ks<1$ (inside the separatrix). This is mainly due to the distribution being highly clustered, so most particles are below the separatrix. Consequently, for most particles, the main contribution to their diffusion comes from resonances with particles at their same frequency. This is represented by the strong yellow line at $\ks<1$.
For the almost homogeneous case, $M_0=0.05$, the particles are not so clustered and so particles with $\ks\ne\kappa$ also contribute, as demonstrated by the presence of other curves in the top panel.

\subsection{Examples of numerical calculations}
\begin{figure}
\begin{centering}
\includegraphics[width=7cm]{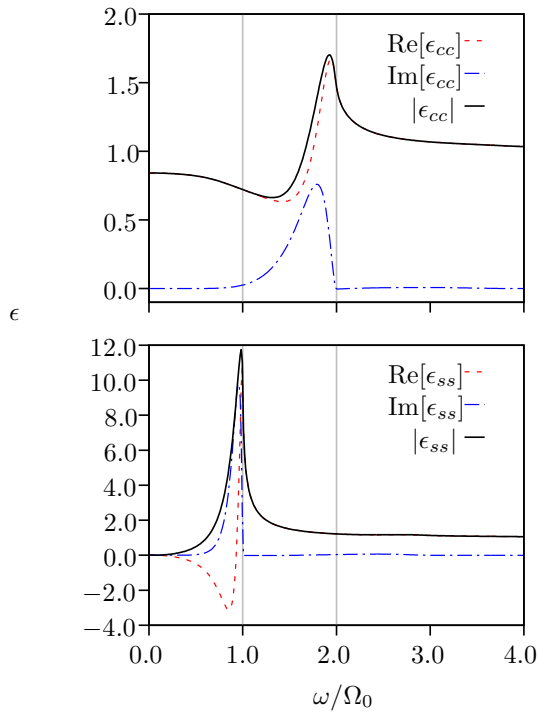}
\caption{Cosine (top) and sine (bottom) components of the dielectric tensor $\epsilon(\omega)$, given by equations~\eqref{eq:hmfepscc} 
and \eqref{eq:hmfepsss}, respectively. The equilibrium parameters are $(u,M_0)=(-0.1,0.728)$. The vertical lines show $\omega=\Omega_0$ and $\omega=2\Omega_0$.}
\label{fig:eps}
\end{centering}
\end{figure}

In this section we show the predictions for the diffusion coefficients both including or neglecting collective effects. Note that, near the separatrix ($\kappa=1$), we do not plot the value of the diffusion coefficient. This is because the calculation becomes numerically unstable in this region. Indeed, the perturbative approach we have used may not be valid~\cite{CarEsc1986,Nei1986} for particles crossing the separatrix. Since it is does not seem to play an important role in the diffusion, we neglect the point $\kappa\approx 1$. First of all, we notice that, as in the homogeneous case ~\cite{bouchet_05},  collective effects are very important in this system. To illustrate this behavior we plot the components of the  dielectric tensor in Figure ~\ref{fig:eps}. We observe a characteristic frequency (materialized by a ``bump'') at a frequency of order $n\Omega_0$, with $n=1$ for sine perturbations and $n=2$ for cosine ones. We observe that collective effects are very important for frequencies $\omega\lesssim n\Omega_0$ in this case, i.e., the modulus of the components of the dielectric tensor is very different from one. Inspecting the kinetic equation \eqref{eq:hmflb}  we see that this implies that for values of $\kappa$ which correspond to these frequencies (which correspond mainly to librating particles) collective effects are important. However, particles with larger frequencies do not present strong collective effects, because they have frequencies $\omega \gg \Omega_0$ for which the components of the dielectric tensor is close to one.

\begin{figure}
\begin{centering}
\includegraphics[width=7.5cm]{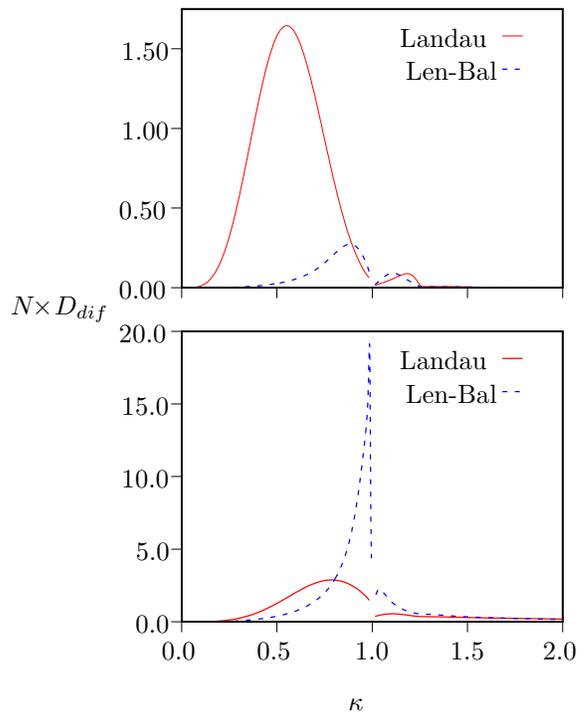}
\caption{Diffusion coefficient $D_{dif}(\kappa)$ for two different equilibrium configurations: $(u,M_0)=(-0.1,0.7285)$ (top) and $(u,M_0)=(0.2475,0.0632)$ (bottom). Solid (red) lines show the diffusion coefficient with collective effects, equations~\eqref{eq:hmfCEdif} and~\eqref{eq:hmfD}, while the dashed (blue) lines show the result without collective effects, equations~\eqref{eq:hmfCEdif} and~\eqref{eq:hmfDbare}. Both curves are cut off near $\kappa=1$ due to numerical instability at the separatrix.\label{fig:hmfCENCEcompare}}
\end{centering}
\end{figure}

This fact is apparent in the computation of the diffusion coefficients for two different magnetizations shown in Figure~\ref{fig:hmfCENCEcompare}. For both small magnetization (i.e. system very close to homogeneity) as well as magnetization closer to $1$, the diffusion coefficients predicted by the Landau equation (no collective effects) and the Lenard-Balescu equation (collective effects) are completely different except, as expected, for $\kappa>1$, which corresponds to particles with frequencies for which the modulus of the components of the dielectric tensor tends to one.

\subsection{Analytical results for highly magnetized states}

It is possible to obtain analytical  expressions for the
diffusion coefficients for highly magnetized configurations. In this case, all the
particles have $\kappa<1$ and it suffices to
perform the sums in the kinetic equations up to $|n|=|n'|=2$ to obtain a good approximation to the dielectric tensor and the diffusion coefficients. This implies that the position of the resonances are $\ks=\kappa$, simply\footnote{Note that in this approximation the flux associated with Eq.~\eqref{eq:LBaa} is zero, and hence $f$ does not vary with time.}. If the system is less magnetized, there are resonances with particles which are outside the separatrix, and in this case it is necessary to solve numerically the resonance condition $n\Omega(\kappa) = n'\Omega(\kappa^*)$. We will study the case in which collective effects are neglected, and then when collective effects are considered for two paradigmatic cases: a core-halo distribution and a Maxwell-Boltzmann distribution. These two distributions can be considered as prototypes of the two classes of distributions which appears after the violent relaxation process. When initial condition leads to a very ``violent'' violent relaxation, it results in a core-halo quasi-equilibrium, while when the initial condition leads to a ``gentle'' violent relaxation, a compact distribution similar to a Gaussian one forms \cite{levin_14}.

\subsubsection{Without collective effects}

When collective effects are neglected,  $\epsilon_{cc}=1$ and $\epsilon_{ss}=1$, a very good approximation is given by taking only the first term of equations~\eqref{eq:hmfCEdif} and \eqref{eq:hmfCEpol} (taking higher terms is straightforward). We
obtain therefore
\bse
\label{coeff-analytic-ncoll}
\begin{align}
D_{dif}(\kappa)&=\frac{4 \pi^8 \kappa^2 \left(1-\kappa^2\right) \text{sech}^4\left(\frac{\pi  K\left(\sqrt{1-\kappa^2}\right)}{2 K\left(\kappa\right)}\right)}{N K\left(\kappa\right)^5 \left(\left(\kappa^2-1\right) K\left(\kappa\right)+E\left(\kappa\right)\right)}f(\kappa)\\
D_{pol}(\kappa)&=\frac{\pi^9 \kappa \left(\kappa^2-1\right) \text{sech}^4\left(\frac{\pi  K\left(\sqrt{1-\kappa^2}\right)}{2 K\left(\kappa\right)}\right)}{2 N \sqrt{M_0} K\left(\kappa\right)^6 \left(\left(\kappa^2-1\right) K\left(\kappa\right)+E\left(\kappa\right)\right)}\frac{\partial f}{\partial \kappa}(\kappa)
\end{align}
\ese
If $M_0$ is very close to $1$, most of the particles have small $\kappa$. It is possible to expand equations~\eqref{coeff-analytic-ncoll} around $\kappa=0$, giving the simple results:
\bse
\begin{align}
\label{coeff-analytic-ncoll-exp}
D_{dif}(\kappa)&= \frac{1}{N}\left(32 \pi^2\kappa^4 + \mathcal O(\kappa^6)\right) f(\kappa)\\
D_{pol}(\kappa)&=\frac{1}{N\sqrt{M_0}}\left(8 \pi ^2 \kappa^3+ \mathcal O(\kappa^5)\right)\frac{\partial f}{\partial \kappa}(\kappa).
\end{align}
\ese

\subsubsection{With collective effects}

We will first consider the core-halo distribution. It can be modeled by the sum of two step functions
\beq\label{eq:hmffch2}
f_{ch}(\kappa)=\eta_1\Theta\left[\mu_1-h\right] + \eta_2 \Theta\left[\mu_2-h\right],
\eeq
where we have assumed that $\mu_1$ and $\mu_2$ corresponds to the energy of particles which are inside the separatrix. Using the definition of $h=M_0(2\kappa^2-1)$ we can express equation~\eqref{eq:hmffch2} as a function of $\kappa$
\beq\label{eq:hmffch2k}
f_{ch}(\kappa)=\eta_1\Theta\left[ 2M_0(\kappa_1^2-\kappa^2) \right] + \eta_2\Theta\left[2M_0(\kappa_2^2-\kappa^2) \right],
\eeq
where $\kappa_i=\sqrt{\mu_i/M_0+1}$ and $\kappa_1<1$ and $\kappa_2<1$.

Computing the dielectric tensor is straightforward because the derivative of $f_{ch}$ about $\kappa$ involves Dirac delta functions:
\beq
\label{partial-ch}
\frac{\partial f_{ch}}{\dd \kappa}=-2 \kappa M_0 \left\{ \eta_1  \delta\! \left[M_0(\kappa_1^2-\kappa^2)\right]+\eta_2  \delta\! \left[M_0(\kappa_2^2-\kappa^2)\right]\right\}
\eeq
The dielectric tensor is purely real, and it can be calculated inserting equation~\eqref{eq:hmffch2k} in equations~\eqref{eq:hmfepscc} and \eqref{eq:hmfepsss}:
\begin{eqnarray}
\label{eq:hmfepscc/ss-ch}
\nonumber
\epsilon_{cc/ss}(\omega)&=&1+2\pi\sum_{\ell=-\infty}^{\infty}\left\{\frac{g_\ell^{cc/ss}(\kappa_1)}{\Om(\kappa_1)-\omega/\ell}+\frac{g_\ell^{cc/ss}(\kappa_2)}{\Om(\kappa_2)-\omega/\ell}\right\}\\
&&+(\omega\to-\omega),
\end{eqnarray}
where $(\omega\to-\omega)$ means to sum the same expression with $\omega$ replaced by $-\omega$. Using equations~\eqref{eq:hmfCEdif} and \eqref{eq:hmfCEpol} with equation~\eqref{eq:hmffch2k} and  $\kappa^*=\kappa$, it is straightforward to compute the diffusion coefficients.

\begin{figure*}
\begin{centering}
\includegraphics[width=18cm]{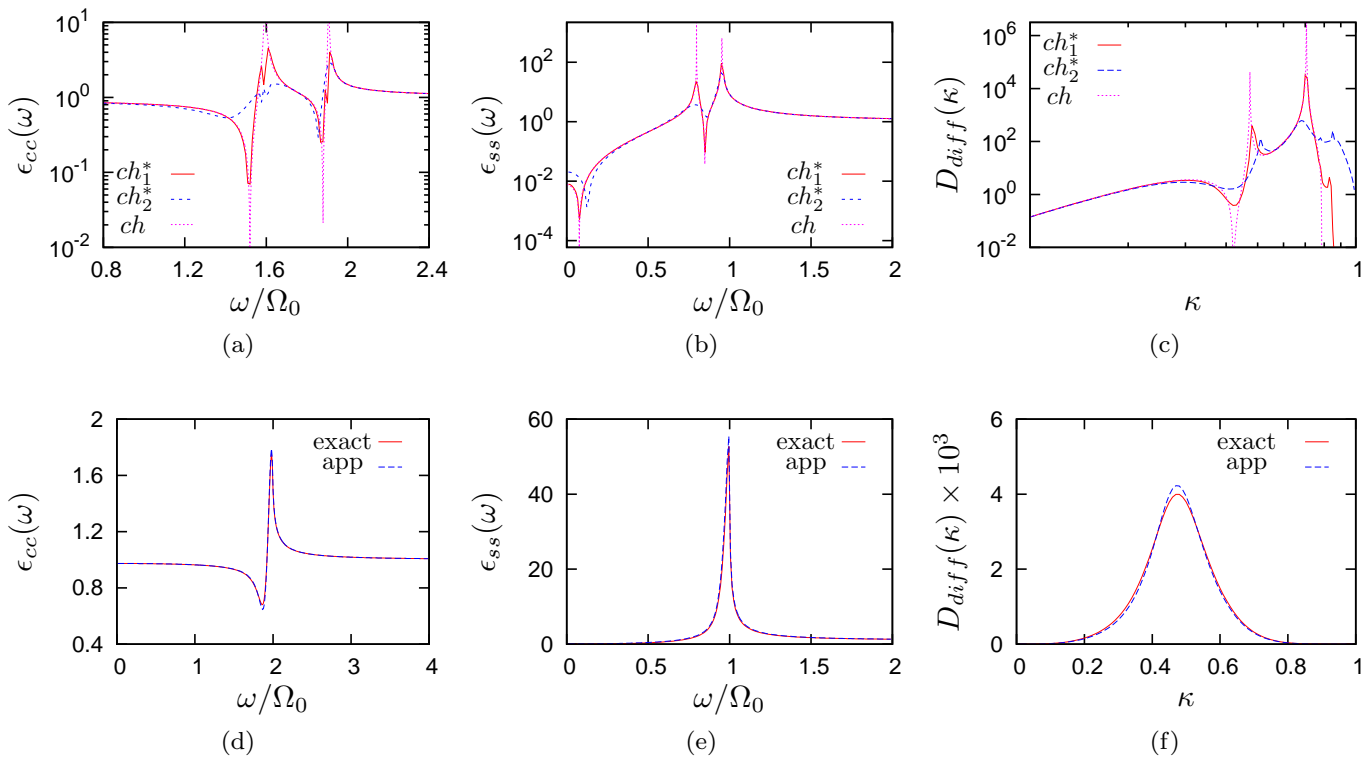}
\caption{\label{M0.8} $(a)$ -- $(c)$ comparison of the approximate expressions \eqref{eps_cc_ap}, \eqref{eps_ss_ap} and the diffusion coefficient for a core-halo system (see text for details), and $(d)$ -- $(f)$ the same quantities at Maxwell-Boltzmann equilibrium for magnetization $M_0=0.95$.}
\end{centering}
\end{figure*}

It is interesting to compare the diffusion coefficients for an idealized core-halo distribution \eqref{eq:hmffch2k} with a more realistic smoother version of it, which is the kind of distribution we simulated (see Section~\ref{subsec:MD}):
\beq\label{eq:hmffch}
f_{ch_i^*}(h)=\frac{\eta_1}{1+\exp[\beta_1(h-\mu_1)]}+\frac{\eta_2}{1+\exp[\beta_2(h-\mu_2)]}.
\eeq
For a given mean energy $u$ and magnetization $M_0$, plus the normalization constraints, three of the six parameters $\eta_1,\eta_2,\beta_1,\beta_2,\mu_1,\mu_2$ are determined. We have chosen the coefficients $\eta_1=0.298$, $\eta_2=0.05$,  $\mu_1=-0.517$ and $\mu_2=0.19$ for $i=1,2$, and $\beta_1=70$, $\beta_2=70$ for $i=1$ and $\beta_1=30$, $\beta_2=10$ for $i=2$. 
As the coefficients $\beta_i$ increase, the step functions become steeper. We observe in the top row of Figure~\ref{M0.8} that for the steeper case $ch_1^*$ the two-step core-halo \eqref{eq:hmffch2k} describes very well both the components of the dielectric tensor and the diffusion coefficient. For the softer case $ch_2^*$, we observe a correct agreement for the components of the dielectric tensor for most of the frequencies. The disagreement is responsible for the differences observed in the diffusion coefficient for some ranges of $\kappa$.

For the case of distributions like the Maxwell-Boltzmann one, the main difficulty consists in computing the dielectric tensor. It is possible to do it analytically for a wide class of functions taking the advantage that if $M_0\to1$, most of the particles have small $\kappa$. We can thus expand in Taylor series the different quantities
which appear in the kinetic equations. We need therefore (valid for
$\kappa\leq 1$):
\bse
\begin{align}
J(\kappa)&=2\sqrt{M_0}\kappa^2 +\mathcal O(\kappa^4) \\
\Omega(\kappa)&=\sqrt{M_0}\left(1-\frac{\kappa^2}{4}+\mathcal O(\kappa^4)\right)\\
c_2(\kappa)&=\frac{\kappa^2}{2}+\mathcal O(\kappa^4)\\
s_1(\kappa)&=-i\kappa+\mathcal O(\kappa^3).
\end{align}
\ese
The components of the dielectric tensor can be approximated as
\bse
\begin{align}
\label{eps_cc_ap_def}
\epsilon_{cc}(\omega)&\simeq 1+\frac{\pi}{2}\int_0^1\dif\kappa\frac{\kappa^4\partial f_{MB}/\partial \kappa}{\sqrt{M_0}\left(1-\frac{\kappa^2}{4}\right)-\omega/2}+(\omega\to-\omega)\\
\label{eps_ss_ap_def}
\epsilon_{ss}(\omega)&\simeq 1+2\pi\int_0^1\dif\kappa\frac{\kappa^2\partial f_{MB}/\partial \kappa}{\sqrt{M_0}\left(1-\frac{\kappa^2}{4}\right)-\omega}+(\omega\to-\omega).
\end{align}
\ese
Taking as the distribution function the  thermal equilibrium one  \eqref{gaussian-def}, the integrals can be expressed in terms of trigonometric and exponential integrals (for the explicit expressions, see Appendix~\ref{app-eps_MB}). Using the approximations \eqref{eps_cc_ap}, \eqref{eps_ss_ap} and the terms  of equations~\eqref{eq:hmfCEdif}
and \eqref{eq:hmfCEpol} corresponding to $n$ and $n'$ taking the values from $-2$ to $+2$ we get, for large $M_0$ a lengthy but analytical
approximation (which we do not explicitly write here) of the diffusion
coefficients which is very accurate for $M_0$ close to one. In the bottom row of Figure~\ref{M0.8}
we show the diffusion coefficients for $M_0=0.95$.

\section{Comparison with simulations}\label{subsec:MD}

The previous subsection presents the application of the kinetic equations to the HMF model. In order to compare those analytical results with the Hamiltonian dynamics of the $N$-body system, we use molecular dynamics, integrating the equations of motion of $N$ particles and tracking their orbits through time.

In order to compare the theoretical results with simulation we adopt the point of view of the Fokker-Planck equation. The idea is to study a test particle evolving in a field composed of the other particles. The effect of the field on the test particle is taken into account by the diffusion and friction coefficients. The mean-field properties of the field evolve adiabatically compared to the timescale of the 
fluctuations which lead to the test particle's relaxation. 
In the case of the HMF model, this means that the field's magnetization is $M=M_0+\delta M$, where $M_0$ evolves very slowly compared to $\delta M$. The test particle's base orbit is thus determined by $M_0$, whereas the fluctuations $\delta M$ drive its relaxation.
The collective effects represent the reaction of the field to its own perturbations, that is,
the field particles are also affected by $\delta M$. If we disregard collective effects, the field
particles should evolve subject only to the mean magnetization $M_0$. Therefore, a possible way
of testing the importance of collective effects in the HMF model is to simulate two types of $N$-body dynamics. 

The first, which we will refer to as ``MD(bath)'', is a dynamics {\em without} collective effects. The system is composed of $N_b$ particles which form a thermal bath and evolve with the adiabatic, static magnetization $M_0$ (corresponding to the smooth potential),
\beq
\ddot{\theta_i}^{b}=-M_0\sin\theta_i,\qquad i=1,\dots,N_b
\eeq
and $N_{tp}$ independent test particles which evolve under the potential due to the oscillating magnetization of the bath particles, 
\beq\label{eq:hmfmdbath}
\begin{aligned}
\ddot{\theta_i}^{tp}=&-M_x^b\sin\theta_i+M_y^b\cos\theta_i,\quad i=N_b\!+\!1,\dots,N_b\!+\!N_{tp}\\
&M_x^b=\frac{1}{N_b}\sum_{i=1}^{N_b}\cos\theta_i,\qquad M_y^b=\frac{1}{N_b}\sum_{i=1}^{N_b}\sin\theta_i.
\end{aligned}
\eeq
The bath particles are set up with any initial positions and velocities corresponding to the Vlasov-stable distribution for which we want to measure the diffusion coefficients, e.g., \eqref{gaussian-def} or \eqref{eq:hmffch}.  We detail the procedure for the former case: the initial particle positions and velocities must be distributed according to
\beq\label{eq:hmfeq}
f_{eq}(\theta,p)=\sqrt{\frac{\beta}{(2\pi)^3}}I_0^{-1}(\beta M_0)\exp\left[-\beta\left(\frac{p^2}{2}-M_0\cos\theta\right)\right].
\eeq
For each $M_0$, $\beta$ must be determined self-consistently by
\beq\label{eq:mageq}
M_0=\frac{I_1(\beta M_0)}{I_0(\beta M_0)}.
\eeq

Second, we simulate  the full $N$-body simulation of the HMF model --- hence {\em with} collective effects --- which we shall refer to as ``MD(full)''. All $N$ particles in the system evolve according to
\beq\label{eq:hmfmdfull}
\begin{aligned}
\ddot{\theta}_i=-M_x\sin\theta_i+M_y\cos\theta_i,\qquad i=1,\dots,N\\
M_x=\frac{1}{N}\sum_{i=1}^N\cos\theta_i,\qquad M_y=\frac{1}{N}\sum_{i=1}^N\sin\theta_i.
\end{aligned}
\eeq
We have seen from the analytical calculations that collective effects are important in the HMF model. Therefore, these two $N$-body methods should result in very different diffusion coefficients. We  measure the diffusion coefficients of test particles as follows: first, we calculate the initial action $J_i(t_0)$ of 
each test particle---or simply each particle, in the case of MD(full)---and separate them accordingly into $L$ bins of size $\Delta J_0$. 
Then, we calculate the mean square variation of $J$ for each $J_0$ as a function of $\Delta t$,
\beq\label{eq:dj2}
\langle\delta J^2\rangle_{\ell}=\frac{1}{N_{\ell}}\sum_{i=1}^{N_{\ell}}[J_i(t_0+\Delta t)-J_0]^2, \qquad \ell=1,\dots,L
\eeq
where the sum, for each bin $\ell$, is over all $N_{\ell}$ particles with $J(t_0)\in [(\ell-1/2) \Delta J_0,(\ell+1/2)\Delta J_0)$.
The diffusion coefficient for a given $J_0$ (or, equivalently, for a given bin $\ell$), is half of the slope of the linear part of the
curve $\langle\delta J^2(\Delta t)\rangle_{\ell}$,
\beq\label{eq:hmfmddif}
D_{dif}^{MD}(J_0)=\frac{\langle\delta J^2\rangle_{\ell}}{2\Delta t}.
\eeq
For some values of $J_0$, care must be taken to calculate the coefficient in the full HMF molecular dynamics: if the magnetization is sufficiently high, there are little to no particles for higher values of $J_0$. 
Therefore, to calculate the coefficient in these regions, we simulate the dynamics of test particles with high $J_0$ that interact with the full HMF.

\begin{figure}
\begin{centering}
\includegraphics[width=8.5cm]{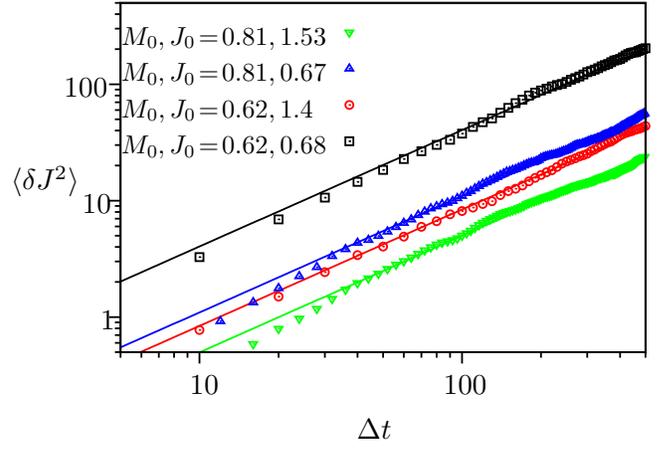}
\caption{Variation of $J^2$, equation \eqref{eq:dj2}, as a function of time, for different values of $J_0$ and different thermal distributions. Points are molecular dynamics results of the regular HMF model and lines are linear fits. For longer times, the diffusion becomes sub-linear.\label{fig:dj2t}}
\end{centering}
\end{figure}

Examples of the linear fit are shown in Figure \ref{fig:dj2t}, for two values of $J_0$. Typically, the fit is done over a time 
range of $t\in[100,500]$, although this may vary depending on the value of $J_0$ and $M_0$. On average, choosing different time ranges does not greatly affect the outcome.
For the fits, we took averages of $\langle\delta J^2(\Delta t)\rangle_{\ell}$ over many time intervals of the dynamics, that is, for many values of $t_0$. Typically, we used $100$ intervals. 

\begin{figure*}
\begin{centering}
\includegraphics[width=16cm]{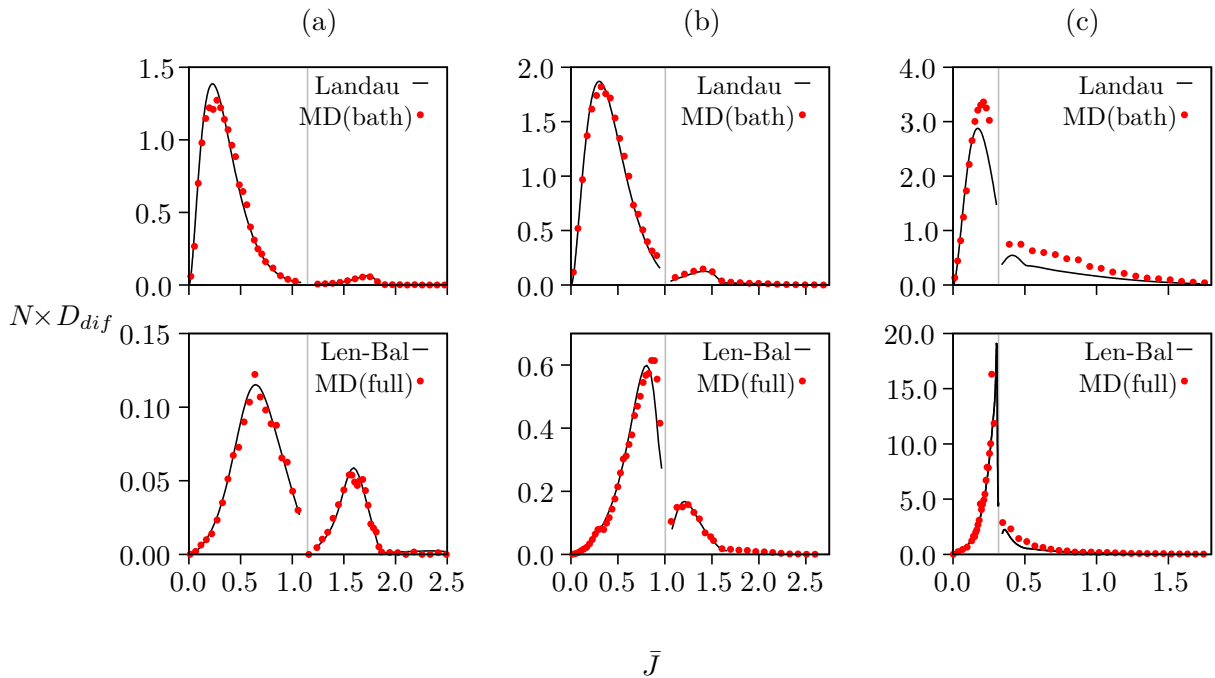}
\caption{\label{fig:hmfdiff}Diffusion coefficients calculated by molecular dynamics, equation \eqref{eq:hmfmddif}, compared to the theoretical results,
 for an equilibrium distribution with parameters (a) $(u,M_0)=(-0.2,0.816)$, (b) $(u,M_0)=(0.0,0.622)$ and (c) $(u,M_0)=(0.2475,0.06)$.
On the bottom, MD simulations without collective effects, with the prediction of the Landau equation \eqref{eq:hmfCEdif}.
On the top, MD simulations with collective effects with the theoretical curve predicted by the Lenard-Balescu (Len-Bal) equation, using condition \eqref{cond-nocoll},
and the molecular dynamics given by the regular HMF model -- MD(full). The gray vertical line represents the separatrix.}
\end{centering}
\end{figure*}

In Figure \ref{fig:hmfdiff}, we compare the molecular dynamics results with the 
kinetic theory diffusion coefficients for systems in thermal baths\footnote{For clarity, in the plots of the diffusion coefficients in which the abscissa is the action, we use instead a rescaled action $\bar{J}$,
\begin{equation*}
\bar{J}=\begin{cases}
J/2&\kappa<1\\
J&\kappa>1.
\end{cases}
\end{equation*}}

The top panels show the case without collective effects
---MD(bath)---  and the Landau diffusion coefficient calculated with \eqref{eq:hmfCEdif} and \eqref{cond-nocoll}, while the bottom panels show the case with collective effects
---MD(full)--- and the Lenard-Balescu diffusion coefficient \eqref{eq:hmfCEdif}. Each kind of simulation has been performed with $N=500000$ particles, except for the lowest magnetization case, which was performed with $N=1000000$.
We see that for magnetizations not close to zero ---panels (a) and (b)--- the MD fit matches very well the result from the corresponding kinetic equation. In the case of magnetization close to zero ---panel (c)--- the match is only reasonably good. This can be explained because in this case the linear diffusion regime is very short and consequently the fluctuations larger.

\begin{figure}
\begin{centering}
\includegraphics[width=8cm]{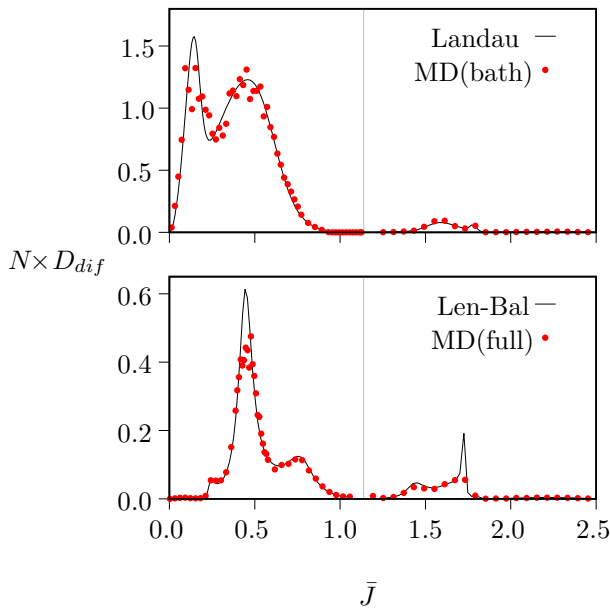}
\caption{Diffusion coefficients for a system in a ``core-halo'' type distribution, given by equation~\eqref{eq:hmffch}.
On the top, without collective
effects: simulation of test particles interacting with the distribution ---MD(bath)--- and the theoretical curve (Landau). The gray vertical line represents the separatrix.
On the bottom, MD simulation results of the regular HMF ---MD(full)--- with the theoretical curve with collective effects (Len-Bal). 
The parameters for the distribution are $\beta_1=30$, $\beta_2=10$, $\eta_1=0.298$, $\eta_2=0.051$, $\mu_1=-0.517$, and $\mu_2=0.19$, which gives $M_0=0.8$. 
\label{fig:hmfch}}
\end{centering}
\end{figure}

We test also the theoretical results for a core-halo distribution $ch_2^*$ equation~\eqref{eq:hmffch}. For both without collective effects (top) and with collective effects (bottom), the results match very well, see Figure~\ref{fig:hmfch}.

\section{Conclusion}

In this paper we have studied the diffusion coefficients  corresponding the collisional relaxation in the inhomogeneous HMF model. To perform these calculations we have used the Landau and the Lenard-Balescu equations expressed in action-angle variables. We have described precisely how to perform the calculations and showed that the diffusion coefficients can be easily computed in a very reduced computer time with high precision. Moreover, we have given analytical expressions for the dielectric tensor and  the diffusion coefficients for systems with magnetization close to one, which agree very well with the exact ones. 

One of the conclusions of the paper is that, for the cases for which we have calculated the diffusion coefficients, collective effects are very important in the dynamics independently of how much the system is clustered (i.e. magnetized). We note that this is also the case in the homogeneous case \cite{bouchet_05}.

We have also studied which particles ``talk to each other'' in the collisional relaxation process. For highly clustered systems (i.e. magnetization close to one), the contribution of the relaxation of a given particle comes almost exclusively from particles in the same orbit (i.e. with the same $\kappa$). This is a similar behavior than in the homogeneous case, for which it is simple to show that for any long-range  one-dimensional system the contribution for the relaxation comes from particles with the same velocity \cite{Cha2012}. As the system becomes less clustered, the situation becomes more complicated, and particles in different orbits start to ``interact'' one with the other (see Figure~\ref{fig:hmfcontribs}). 

In order to test the theoretical predictions we have computed numerically the diffusion coefficients using molecular dynamics simulations. To check our calculations when the collective effects are neglected, we have set up a simple method to perform simulations in which collective effects are absent. We have found a very good agreement between the theoretical calculations and the simulations both for the dynamics with and without collective effects. We have performed these tests for baths at Maxwell-Boltzmann equilibrium as well as out-of-equilibrium (core-halo distributions).

The next natural step of this work is to use the diffusion coefficients to compute the whole evolution of the HMF model up to thermalisation. With the methods developed in the paper it is a relatively simple task to compute the evolution with the Landau or the  Lenard-Balescu equation. The magnetization should be computed self-consistently at each time step and then the diffusion coefficient. We stress that the evolution of equation~\eqref{eq:hmflb} could present interesting features because it is non-linear.
This subject will be presented in a forthcoming paper.

We note also that the analytical expressions for the dielectric tensor can be used to study analytically the stability and the mean-field evolution of the HMF model for highly clustered states, computing in an appropriate but straightforward way the pole contributions to the dielectric tensor (see \cite{BarOli2010} for a detailed study on the subject).

The extension of our calculations to more complicated interactions, e.g. one-dimensional gravity, is in principle feasible. There are however two complications to the calculations compared to the HMF model: first, the bi-orthogonal basis is not constituted by only two functions, but by a infinite number of them. There is however the hope that with a suitable choice of family of functions for a given shape of the QSS a reduced number of elements of the basis is sufficient to obtain a good accuracy in the calculations, similarly to the case studied in \cite{fouvry_15,fouvry_15b}. Second, we do not expect to have an analytical expression for the Fourier transform of the angle of the element of the basis (equations~\eqref{eq:hmfft1}). These calculations should be performed numerically, which is feasible with a modest computer.

\begin{acknowledgments}

The authors warmly thank J. Barr\'e, D. Chiron, T.M.R. Filho, J.B. Fouvry, A. Galligo, D. M\'etivier, and T.N. Teles for interesting discussions. This work was partially funded by the Brazilian agencies CNPq and CAPES. This work was granted access to the HPC and visualization resources of the Centre de Calcul Interactif hosted by Universit\'e Nice Sophia Antipolis, the Mesocentre SIGAMM machine hosted by the Observatoire de la C\^ote d'Azur, and the Center of Computational Physics CFCIF of the Universidade Federal do Rio Grande do Sul.

\end{acknowledgments}

\appendix
\section{Action-angle variables of the pendula}
\label{ap:aa}
In this appendix, we present action-angle variables for a pendulum with the
Hamiltonian
\beq
h(\theta,p)=\frac{p^2}{2}-M_0\cos\theta,
\eeq
using the same conventions as references \cite{BarOli2010} and \cite{BarMet2016}.
The action $J$ is given by
\beq
J=\frac{1}{2\pi}\oint p\dif\theta.
\eeq
If the energy $h$ is greater than the magnetization $M_0$, the orbit is rotating: its momentum will never reach zero. 
In such cases, the integration over $\theta$ will only go from $-\pi$ to $\pi$, 
for positive momentum, or $\pi$ to $-\pi$, for negative momentum. 
For librating orbits, which have energy $h$ less than the magnetization $M_0$, 
the orbit completes a loop in phase space (see Figure \ref{fig:hmforbits} in the main text), reaching zero momentum at the extreme value of $\theta$, $\pm\theta_m$.
The integration starts with positive momentum at $-\theta_m$, goes to $\theta_m$, and then back to $-\theta_m$ with negative momentum. The action is thus given by
\beq
J=\frac{1}{2\pi}
\begin{cases}\label{eq:hmfaction1}
2\int_{-\theta_m}^{\theta_m}\sqrt{2(h+M_0\cos\theta)}\dif\theta&\qquad h<M_0,\\
\int_{-\pi}^{\pi}\sqrt{2(h+M_0\cos\theta)}\dif\theta&\qquad h>M_0.
\end{cases}
\eeq
Using the transformation $x=\theta/2$ and $\cos\theta=1-2\sin^2(\theta/2)$, equation \eqref{eq:hmfaction1} can be written as
\begin{equation}\label{eq:hmfaction2}
J=\frac{4\sqrt{M_0}}{\pi}
\begin{cases}
2\int_{0}^{\frac{\theta_m}{2}}\sqrt{\kappa^2-\sin^2x}\dif x&\qquad\kappa<1,\\
\kappa\int_0^{\frac{\pi}{2}}\sqrt{1-\frac{1}{\kappa^2}\sin^2x}\dif x&\qquad\kappa>1,
\end{cases}
\end{equation}
where
\beq\label{eq:hmfkappaapend}
\kappa=\sqrt{\frac{h+M_0}{2M_0}}.
\eeq
and $\theta_m=2\arcsin(\kappa)$.
For $\kappa>1$, the integral in equation \eqref{eq:hmfaction2} is the complete Legendre elliptic integral of the second kind $\E(1/\kappa)=\E(\pi/2,1/\kappa)$, where
\beq\label{eq:clei2}
\E(\phi,k)=\int_0^{\phi}\sqrt{1-k^2\sin^2\theta}\dif\theta,\qquad k<1.
\eeq
For $\kappa<1$, switching variables with $\sin\theta=\kappa\sin x$, the corresponding
integral in equation \eqref{eq:hmfaction2} becomes
\beq
\int_0^{\theta_m/2}\sqrt{\kappa^2-\sin^2x}\dif x=\E(\kappa)-(1-\kappa^2)\K(\kappa)
\eeq
where $\K(k)$ is the complete elliptic integral of the first kind,
\beq
\K(k)=\int_0^{\pi/2}\frac{\dif\theta}{\sqrt{1-k^2\sin^2\theta}}.
\eeq
Therefore, the action is
\beq
J=
\begin{cases}
\frac{8\sqrt{M_0}}{\pi}\left[\E(\kappa)-(1-\kappa^2)\K(\kappa)\right],&\qquad\kappa<1,\\
\frac{4\sqrt{M_0}}{\pi}\kappa\E\left(\frac{1}{\kappa}\right),&\qquad\kappa>1.
\end{cases}
\eeq

The angle variables, $w$, satisfy \cite{lichtenberg2010regular}
\beq
w=\Omega t
\eeq
where $\Om=\partial h/\partial J$ is the angular frequency and $t$ is the time of the pendulum at position $\theta$,
\beq
t=\int_0^{\theta}\frac{\dif\theta'}{\sqrt{2(h+M_0\cos\theta')}}.
\eeq
Integrating $\int \dif t=\int \dif \theta/p(\theta,\kappa)$ gives
\beq\label{eq:t}
t(\theta,\kappa)=\frac{1}{\sqrt{M_0}}
\begin{cases}
\F\left(\phi,\kappa\right)&\qquad \kappa<1,\,\,p>0,\\
2\K(\kappa)-\F\left(\phi,\kappa\right)&\qquad \kappa<1,\,\,p<0,\\
\frac{1}{\kappa}\F\left(\frac{\theta}{2},\frac{1}{\kappa}\right)&\qquad \kappa>1,\,\,p>0,\\
\frac{1}{\kappa}\F\left(\frac{\theta}{2},\frac{1}{\kappa}\right)&\qquad\kappa>1,\,\,p<0,
\end{cases}
\eeq
where $\phi=\arcsin\left(\frac{1}{\kappa}\sin\frac{\theta}{2}\right)$.
Multiplying by $\Om(\kappa)$ as given by equation \eqref{eq:hmfomega}, we find the angle variables
\beq\label{eq:w}
w=\pi
\begin{cases}
\frac{\F\left(\phi,\kappa\right)}{2\K(\kappa)}&\qquad \kappa<1,\,\,p>0,\\
1-\frac{\F\left(\phi,\kappa\right)}{2\K(\kappa)}&\qquad \kappa<1,\,\,p<0,\\
\frac{\F\left(\frac{\theta}{2},\frac{1}{\kappa}\right)}{\K\left(\frac{1}{\kappa}\right)}&\qquad\kappa>1,\,\,p>0,\\
\frac{-\F\left(\frac{\theta}{2},\frac{1}{\kappa}\right)}{\K\left(\frac{1}{\kappa}\right)}&\qquad\kappa>1,\,\,p<0.
\end{cases}
\eeq

\section{Elliptic identities for Fourier transforms}\label{ap:ellip}

In this appendix, we show how to obtain the expressions for the Fourier transforms of the orthogonal components of the potential, proportional to $c_n(\kappa)$ and $s_n(\kappa)$ (equation \eqref{eq:hmfft1}), as obtained in reference \cite{BarMet2016}.
First, we must find $\cos[\theta(w,\kappa)]$ and $\sin[\theta(w,\kappa)]$ as functions of $w$ and $\kappa$ directly. 
These can be obtained from the angle variable \eqref{eq:w}, which depends on $\theta$ through incomplete elliptic integrals \cite{BarOli2010}. 
For the incomplete elliptic integral of the first kind $\F(\alpha,k)$, $\alpha$ can be expressed in terms of the Jacobi elliptic functions $\sn(u,k)$, $\cn(u,k)$ and $\dn(u,k)$. In particular, if $\F(\alpha,k)=u$, then $\sin\alpha=\sn(u,k)$.
Applying to equation \eqref{eq:w} gives
\beq\label{eq:cosellip}
\cos[\theta(w,\kappa)]=
\begin{cases}
1-2\kappa^2\sn^2\left(\frac{2\K(\kappa)w}{\pi},\kappa\right)&\qquad \kappa<1,\\
1-2\sn^2\left(\frac{\K(1/\kappa)w}{\pi},1/\kappa\right)&\qquad \kappa>1,
\end{cases}
\eeq
and
\begin{align}\label{eq:sinellip}
\sin[\theta&(w,\kappa)]=\nonumber\\
&
\begin{cases}
2\kappa\sn\!\left(\frac{2\K(\kappa)w}{\pi},\kappa\right)\!\dn\!\left(\frac{2\K(\kappa)w}{\pi},\kappa\right)&\quad\kappa<1,\\
2\sn\!\left(\frac{\K(\frac{1}{\kappa})w}{\pi},\frac{1}{\kappa}\right)\!\cn\!\left(\frac{\K(\frac{1}{\kappa})w}{\pi},\frac{1}{\kappa}\right)&\quad\kappa>1,\,\,p>1,\\
\!-2\sn\!\left(\frac{\K(\frac{1}{\kappa})w}{\pi},\frac{1}{\kappa}\right)\!\cn\!\left(\frac{\K(\frac{1}{\kappa})w}{\pi},\frac{1}{\kappa}\right)&\quad\kappa>1,\,\,p<1,
\end{cases}
\end{align}
where the properties $\sn^2(u,k)+\cn^2(u,k)=1$ and $\dn(u,k)=\sqrt{1-k^2\sn^2(u,k)}$ were used.
Finally, \eqref{eq:cosellip} and \eqref{eq:sinellip} can be expressed in terms of the following expansions involving the elliptic functions \cite{Mil2002},
\beq\label{eq:milnsn2}
\sn^2(u,k)\!=\!\frac{\K(k)\!-\!\E(k)}{k^2\K(k)}-\frac{2\pi^2}{k^2\K(k)^2}
\!\sum_{n=1}^{\infty}\frac{n\q(k)^n}{1\!-\!\q(k)^{2n}}\!\cos\frac{\pi nu}{\K(k)},
\eeq
\beq\label{eq:milnsndn}
\sn(u,k)\!\dn(u,k)\!=\!\frac{2\pi^2}{k\K(k)^2}\!\sum_{n=1}^{\infty}\!
\frac{(n\!-\!\frac{1}{2})\q(k)^{n-\frac{1}{2}}}{1\!+\!\q(k)^{2n-1}}\sin\frac{\pi(n\!-\!\frac{1}{2})u}{\K(k)},
\eeq
\beq\label{eq:milnsncn}
\sn(u,k)\!\cn(u,k)\!=\!\frac{2\pi^2}{k^2\!\K(k)^2}\!\sum_{n=1}^{\infty}
\frac{n\q(k)^n}{1+\q(k)^{2n}}\sin\frac{\pi nu}{\K(k)}
\eeq
where $\q(k)=\exp[-\pi\K(\sqrt{1-k^2})/\K(k)]$.

To find $c_n(\kappa)$ and $s_n(\kappa)$, the above expansions should 
be applied in the equations for $\cos[\theta(w,\kappa)]$ and $\sin[\theta(w,\kappa)]$. This gives the results of equations \eqref{eq:hmfcm} and \eqref{eq:hmfsm}.

\section{Dielectric tensor for a Maxwell-Boltzmann distribution for $M_0\to1$}
\label{app-eps_MB}
Taking as the distribution function the  thermal equilibrium one \eqref{gaussian-def}, the components of the dielectric tensor can be approximated as
\begin{eqnarray}
\label{eps_cc_ap}
\nonumber
\epsilon_{cc}(\omega)&\simeq& 1+\frac{\pi}{2}\int_0^1\dif\kappa\frac{\kappa^4\partial f_{MB}/\partial \kappa}{\sqrt{M_0}\left(1-\frac{\kappa^2}{4}\right)-\omega/2}+(\omega\to-\omega)\\\nonumber
&\simeq&1+\frac{16 \pi  \beta C\left(\omega-2 \sqrt{M_0}\right)^2 \alpha_1 \left(\text{Ei}\left(x_1\right)-\text{Ei}\left(x_2\right)\right)}{\sqrt{M_0}}\\\nonumber
&&+\frac{2 \pi  C \left(\alpha_2 \sinh (\beta M_0)-\beta M_0 \cosh (\beta M_0)\right)}{\beta M_0^{3/2}}\\\nonumber
&&+i\frac{16 \pi^2 b C \left(w-2 \sqrt{M_0}\right)^2 \alpha_1 \Theta \left(\sqrt{M_0}-\frac{w}{2} \right)\Theta(\omega)}{\sqrt{M_0}}\\
&&+(\omega\to-\omega).
\end{eqnarray}
\begin{eqnarray}
\label{eps_ss_ap}
\nonumber
\epsilon_{ss}(\omega)&\simeq& 1+2\pi\int_0^1\dif\kappa\frac{\kappa^2\partial f_{MB}/\partial \kappa}{\sqrt{M_0}\left(1-\frac{\kappa^2}{4}\right)-\omega}+(\omega\to-\omega)\\\nonumber
&\simeq& 1+ 64 \pi \frac{\sinh (b M_0)}{\sqrt{M_0}}\\\nonumber
&&-64\pi b \left(\sqrt{M_0}-w\right) \alpha_3 \left(\text{Ei}\left(x_3\right)-\text{Ei}\left(x_4\right)\right)\\\nonumber
&&+i 16 \pi ^3 b C \left(\sqrt{M_0}-w\right) \alpha_3 \Theta \left(\sqrt{M_0}-w\right)\Theta(\omega)\\
&&+(\omega\to-\omega),
\end{eqnarray}
where where $\alpha_1= e^{4 \beta \sqrt{M_0} \omega-7 \beta M_0}$, $\alpha_2=-4 \beta \sqrt{M_0} \omega+9 \beta M_0+1$, $\alpha_3=e^{8 b \sqrt{M_0} w-7 b M_0}$,  $x_1=6 \beta M_0-4 \beta \sqrt{M_0} \omega$  $x_2=8 \beta M_0-4 \beta \sqrt{M_0} \omega$, $x_3=8 b \left(M_0-\sqrt{M_0} w\right)$,  $x_4=6 b M_0-8 b \sqrt{M_0} w$, $\Theta(x)$ is the Heaviside step function and $(\omega\to-\omega)$ to sum to the expressions written the same with $\omega$ replaced by $-\omega$.

\end{document}